# Enhanced Magnetization and Conductivity in NiFe$_2$O$_4$

Ching Cheng*

*Department of Physics, National Cheng Kung University, Tainan 70101, Taiwan*

Different configurations of magnetic orders and cation distributions for NiFe$_2$O$_4$ are studied by the density functional based methods with the possible inclusion of the on-site Coulomb interaction U.  The lowest energy state is an inverse spinel structure of tetragonal space group P4$_3$22 with Ni locating at the octahedral site (B site) while the Fe distributing equally at the tetrahedral (A site) and octahedral site and with antiparallel moments between cations at A and B sites.  However, a quadruple enhanced magnetization is obtained for a structure with the same cation distribution as that of the lowest energy state, i.e. still an inverse spinel structure, but with antiparallel moments between Ni and Fe cations.   Its energy is higher than that of the ground state by about 0.4eV/fomula.  Both phases are insulating.  Another phase with enhanced magnetization is a normal spinel structure with Ni and Fe locating at A and B site respectively, i.e. corresponding to a structure of cation inversion.  Its electronic structure displays a conductive and nearly half-metallic behavior.  These two phases could be the possible candidates for the experimentally observed NiFe$_2$O$_4$ thin films possessing considerably enhanced magnetization but different electric properties, i.e. one insulating and the other conductive with spin-polarized current.

*ccheng@mail.ncku.edu.tw

PACS number(s): 75.50Gg, 71.15.Mb,75.70.-i, 75.75.-c

Bulk nickel ferrite, $NiFe_2O_4$ (NFO), is an insulator (or semiconductor) with a room-temperature resistivity of ~1 kΩ · cm and shows soft ferrimagnetic order below 850 K with a relatively low magnetization of 2 $\mu_B$ per formula unit (2$\mu_B$/fu), i.e. about 300 emu . cm$^{-3}$ [1]. It is considered to have a completely inverse spinel structure in its bulk form. Recent advances on NFO nanometer films found enhanced magnetic moments and possible tuning of their electronic properties from insulating to conductive state possessing highly spin-polarized current, which hence can be integrated for potential applications in spintronics [2~5]. The enhanced magnetization and conductive property is commonly attributed to cation inversion. In this study, we present a thorough study on the magnetic and electronic properties of NFO through exploring various magnetic configurations and cation distributions of the material using density functional based methods. We shall show that the energetically most probable phase exhibiting enhanced magnetization is an inverse spinel phase with insulating property while the metallic property only exists in the normal spinel phases, i.e. due to the cation inversion.

The structure of spinel ferrites ($AB_2O_4$) are constructed by filling one-eighth of the tetrahedral sites and half of the octahedral sites (denoted as A and B site, respectively, hereafter) in the fcc sublattice of oxygen. In the inverse spinel structure of the bulk NFO, the Ni cations occupy the B site while the Fe cations distribute equally between the A and B sites. A normal spinel, which is considered as a structure with cation inversion for NFO, corresponds to the structure with the A sites being occupied by the Ni cations and all the B sites by the Fe cations. Throughout this study, the cubic unit cell consisting of eight formulas is used in order to examine the chemical and magnetic interactions between cations. In addition to the distinction between the inverse and normal spinel structures, there are several possible distributions for cations at B sites (divided into two groups and denoted as the B1 and B2 sites) as well as different magnetic configurations for the cations at A, B1 and B2 sites [6]. The cation distribution at B site for the structures discussed in this study corresponds to the SC3 distribution in the reference [6] as its energy is always found being the lowest among the four distributions (SC1~SC4) having the same magnetic configurations. The phases are denoted as, e.g. Ipnn (Npnn), which corresponds to the inverse (normal) spinel structure with cations at A, B1 and B2 sites occupied by Fe, Fe and Ni (Ni, Fe and Fe) ions respectively while the magnetic moments are anti-parallel between A and B sites and parallel among B sites. Table I lists all the phases considered in this study.

Table I: The notation with their corresponding cation distributions and magnetic configurations used for the phases considered in this study. The A and B (B1 and B2) sites are the tetrahedral and octahedral sites of cations in the spinel structure. The magnetic configurations are represented by the parallel (++ or - - ) or antiparallel (+ - ) magnetic moments between the cations.

| site | Inverse phases | | | | | Normal phases | | | |
|------|--------|------|------|------|------|--------|------|------|------|
|      | cation | Ipnn | Ipnp | Ippn | Ippp | cation | Npnn | Nppn | Nppp |
| A    | Fe     | +    | +    | +    | +    | Ni     | +    | +    | +    |
| B1   | Fe     | -    | -    | +    | +    | Fe     | -    | +    | +    |
| B2   | Ni     | -    | +    | -    | +    | Fe     | -    | -    | +    |

In the present study, all the electronic calculations are based on the spin polarized density functional theory [7,8] implemented in the plane-wave-based Vienna ab initio simulation program (VASP) which was developed at the Institute fur Material Physik of the Universitat Wien [9]. The generalized gradient approximation (GGA) developed by Perdew, Burke, and Ernzerhof is used for the exchange-correlation energy functional [10]. In transition metals, the on-site Coulomb repulsion is included to account for the strong correlation of the d-orbital electrons [11]. The values of U and J for Fe are taken from the previous studies as 4.5 eV and 0.89 eV [11,12], respectively. For Ni, the values of 2eV, 4eV, and 6eV of U, all with J=1.0 eV, are included in order to study the trend of the U effect on energetics, magnetization and electronic property [13]. The interaction between ions and valence electrons is described by the projector augmented wave (PAW) method [14] which was implemented by Kresse and Joubert [15]. The valence electrons included are 9, 8, and 6 for Ni, Fe, and oxygen respectively. The energy cutoff that determines the number of the plane waves is 500eV and the k-point sampling according to Monkhorst-Pack [16] is (6 6 6) for the 8-formula cubic cell. Previous study for $ZnFe_2O_4$ has shown that higher energy cutoffs and denser k-point sets would change the exchange-energy values by no more than 1 meV [6]. Atomic relaxation and volume relaxation are implemented in turn until the atomic forces are smaller than 0.02 eV/Å and the stresses are less than 3 KBar while keeping the cubic unit cell.

The calculated relative energy and magnetic moments of all the studied phases in Table I are plotted in Fig.1. In addition to the case without U effect (denoted as U0 case hereafter), four different U cases are considered here. They all include the on-site U for Fe, but are accompanied with four different U values for Ni, i.e. one without U (U=J=0) and the other three with U=2, 4 and 6 eV (all with J=1eV). No matter which (and whether or not) on-site Coulomb interactions are included in the calculations, the most stable state is always the Ipnn and its magnetization, i.e. $2\mu_B$/fu, is consistent with the experimental value for the bulk NFO [1]. The antiparallel moments of the Fe ions at A and B sites are cancelled and the net moment of the phase comes from the moment of the Ni ions at B site. The theoretical lattice constant is 8.40Å, with a less than 1% difference between results from the different U cases, which is compared closely to the experimental lattice constant of 8.34 Å [1]. The space group of this phase, taking into account of both the cation distribution and magnetic configuration, is $P4_322$ of the tetragonal space group. Note that the previously examined structure [17-21] is commonly the SC1 of reference [6] which corresponds to the Imma in the orthorhombic space group. A recent experimental article has reported evidence for a short-range B-site order in NiFe$_2$O$_4$ with possible tetragonal $P4_122/P4_322$ symmetry, though the phase with Imma space group was not ruled out [22]. Our calculations always lead to an energetic preference of $P4_322$ over Imma, though by various amounts in different U cases, all within the range between 10 and 20 meV/fu. The different physical properties between these two phases also show up in the electronic structure. For the U0 case, the SC1 phase (Imma symmetry) is a conductor while the SC3 phase ($P4_322$ symmetry) has a band gap of 0.20eV. The electronic property will be discussed in details later.

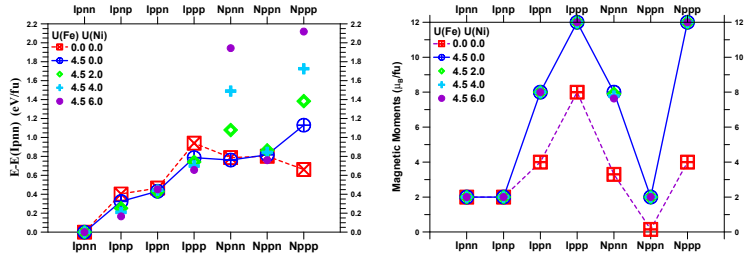

Figure 1: The calculated energies (a), relative to the lowest-energy phase of Ipnn, and the saturation moment (b) of the phases in Table I.

Of the four inverse spinel phases studied here which have the same cation distribution but different magnetic configurations, two phases, i.e. Ippn and Ippp, have magnetizations much more enhanced than the lowest-energy Ipnn phase. The Ippn phase has a magnetization which is either twice or four times that of the Ipnn, depending on whether the on-site Coulomb interaction is included or not. In the U0 case, the anti-parallel moments of the B-site Fe and B-site Ni (both around 2 $\mu_B$) are cancelled and the magnetization arises solely from the A-site Fe which is 4 $\mu_B$/fu. In the non-zero U cases, both the parallel moment of the A-site and B-site Fe contribute 5 $\mu_B$/Fe, together with the antiparallel moment of Ni (2 $\mu_B$/Ni) they lead to a total of 8 $\mu_B$/fu. The energies of Ippn are higher than those of Ipnn for all the U cases by around 0.4 eV/fu. Although there exists the phase Ipnp whose energy is between the lowest-energy Ipnn phase and the enhanced-magnetization Ippn phase, the Ipnp has the same magnetization as that of Ipnn, i.e. 2 $\mu_B$/fu. Consequently the Ippn phase is the energetically most likely phase in the bulk form which might contribute to the much enhanced magnetization in the thin film or nanoparticles of NFO. This enhanced magnetization is achieved by different magnetic configuration from the lowest-energy Ipnn phase in the inverse spinel structure and accordingly involves no cation inversion, as generally proposed in the previous studies [2-5]. The Ippp phase has an even higher magnetization than that of Ippn, i.e. 8 µ$_B$/Fe and 12 µ$_B$/Fe for the U0 and non-zero U cases respectively. However, it is also higher in energy than the Ippn phase by about 0.2~0.4 eV/fu, depending on the U effect.

Of the three magnetic configurations in the normal spinel structure, both the Npnn and Nppp phases have much higher magnetizations than that of the ground state in all the non-zero U cases, with a saturation moment of 8 µ$_B$/fu and 12 µ$_B$/fu respectively. In the U0 case, both phases have a saturation moment of 4 µ$_B$/fu. The energies of Npnn and Nppp, when relative to those of the lowest-energy Ipnn phase, display a much stronger U (of Ni) dependence than the Nppn and the inverse spinel phases. We shall see that this can be attributed to the increase of the interactions among Fe ions at B sites due to the increasing on-site Coulomb repulsion of Ni ions at A sites.

Using the Heisenberg model $H = -1/2 \sum_n \sum_i J_n S_i S_{i+n}$ to describe the magnetic interactions between the cations, we can obtain the three nearest-neighbor interactions for the inverse spinel phases, i.e. $J1_{AB1}$ between the A-site Fe and B1-site Fe, $J1_{AB2}$ between the A-site Fe and B2-site Ni, and $J1_{B1B2}$ between the B1-site Fe and B2-site Ni, from the calculated energies of the four inverse phases of Ipnn, Ipnp, Ippn and Ippp. Note that the cation distributions for the four inverse phases considered here are identical and their differences are solely from the different magnetic configurations. The results are listed in Table II. All the three interactions are found antiferromagnetic. However, the interactions between the B-site cations are one order in magnitude smaller than those between the A-site and B-site cations, no matter whether they are between the A-site Fe and B-site Fe or between the A-site Fe and B-site Ni. These results are the main reason why the most stable phase has the magnetic configuration of Ipnn, a configuration sacrifices the weak antiferromagnetic interaction among B-site cations for the strong antiferromagnetic interaction between A-site and B-site cations.

Table II: The nearest-neighbor exchange interactions (in unit of meV) for the inverse spinel structures at different U cases. Negative sign indicates an antiferromagnetic interaction.

| U(Fe) U(Ni) | $J1_{AB1(FeFe)}$ | $J1_{AB2(FeNi)}$ | $J1_{B1B2(FeNi)}$ |
|---|---|---|---|
| 0.0  0.0 | -41.8 | -36.6 | -4.4 |
| 4.5  0.0 | -37.3 | -28.3 | -2.0 |
| 4.5  2.0 | -38.3 | -23.6 | -3.8 |
| 4.5  4.0 | -38.9 | -19.4 | -2.9 |
| 4.5  6.0 | -39.3 | -15.5 | -2.4 |

The nearest-neighbor interactions for the normal spinel structures in the non-zero U cases are also found to be antiferromagnetic for both the AB and BB interactions (Table

III). As the on-site U effect of Ni increases, the AB interaction goes down as expected. On the other hand, the BB interaction between B-site Fe ions was found increasing rapidly. This demonstrates that the on-site Coulomb repulsion of the A-site Ni has strong effect on the interaction among B-site Fe ions which is not apparent from inspecting the magnitudes of $J1_{AB}$ alone.

Table III: The nearest-neighbor exchange interactions (in unit of meV) for the normal spinel structures at different U cases.

| U(Fe) U(Ni) | $J1_{AB(NiFe)}$ | $J1_{BB(FeFe)}$ |
|---|---|---|
| 0.0 0.0 | +5.2 | +9.6 |
| 4.5 0.0 | -15.3 | -16.1 |
| 4.5 2.0 | -12.7 | -46.0 |
| 4.5 4.0 | -9.8 | -95.7 |
| 4.5 6.0 | -7.3 | -158.9 |

The calculated electronic properties are summarized in Table IV. In the non-zero U cases, all the inverse phases are insulators while the normal Npnn and Nppp phases (the two with much enhanced magnetization) are either half metals or metals. As the bulk NFO has a room-temperature resistivity of ~1 kΩ·cm, the band gap is expected to be about the same order as that of the bulk Si (1.22eV at 0K). This is most closely related to the ground state phase Ipnn at the U cases of U(Fe=4.5,Ni=0) and U(Fe=4.5,Ni=2). The Nppp phase is always found a half metal, independent of the U effects included in the calculations. The Npnn phase is a half metal in the case of U(Fe=4.5,Ni=0). For the cases of U(Fe=4.5,Ni=2) and U(Fe=4.5,Ni=4), the density of states (DOS) is close to a half metal except for a short tail (~0.2eV) of low DOS for the minority-spin component extending below Fermi level. Therefore the Npnn phase is expected to be conductive with high spin-polarized current. Together with the results in energetics and magnetization of Fig. 1, the experimentally observed NFO thin films and nanoparticles with much enhanced magnetization are likely to be due to the presence of the Ippn phase, which is insulating, and the Npnn phase, which is (or nearly) a half metal.

Table IV: The band gaps (in unit of eV) of the studied phases at different U cases. HM and M denotes half metal and metal respectively. The notation of ~HM indicates a nearly half metal property (details in text).

| U(Fe) U(Ni) | Ipnn | Ipnp | Ippn | Ippp | Npnn | Nppn | Nppp |
|---|---|---|---|---|---|---|---|
| 0.0 0.0 | 0.20 | HM | HM | HM | M | M | HM |
| 4.5 0.0 | 1.05 | 0.75 | 0.53 | 0.37 | HM | HM | HM |
| 4.5 2.0 | 1.47 | 1.47 | 0.92 | 0.77 | ~HM | 0.66 | HM |
| 4.5 4.0 | 1.76 | 1.77 | 1.29 | 1.08 | ~HM | 1.46 | HM |
| 4.5 6.0 | 1.92 | 2.02 | 1.33 | 1.17 | M | 1.86 | HM |

The energetics, magnetic and electronic properties of bulk $NiFe_2O_4$ at different magnetic configurations and cation distributions are investigated by the density-functional based methods. It is demonstrated that the ground-state phase has a tetragonal symmetry space group of $P4_322$. The energetically most likely phase which has enhanced magnetization is the Ippn phase which is insulating and an inverse spinel structure. The experimentally observed conducting phase with spin-polarized current probably originates from the normal Npnn phase with cation inversion and an electronic property of nearly half metal.

Acknowledgements:
This work was sponsored by the National Science Council of Taiwan. Part of the computer resources are provided by the NCHC (National Center of High-performance Computing). We also thank the support of NCTS (National Center of Theoretical Sciences) through the CMR (Computational Material Research) focus group.